\pdfoutput=1
\documentclass[12pt,a4paper]{article}

\usepackage{ifthen} 
\newboolean{pdflatex}
\setboolean{pdflatex}{true} 

\newboolean{articletitles}
\setboolean{articletitles}{true} 

\newboolean{uprightparticles}
\setboolean{uprightparticles}{false} 

\def\paperauthors{LHCb collaboration} 
\def\paperasciititle{Observation of Lambdab to Lambdac D(star)0bar K and Lambdab to Lambdac Dsstar decays} 
\def\papertitle{Observation of $\LbLcDzbstK$ and $\LbLcDss$ decays} 
\def\paperkeywords{{High Energy Physics}, {LHCb}} 
\def\papercopyright{\the\year\ CERN for the benefit of the LHCb collaboration} 
\def\paperlicence{CC BY 4.0 licence}
\def\paperlicenceurl{https://creativecommons.org/licenses/by/4.0/}

\usepackage[top=1in, bottom=1.25in, left=1in, right=1in]{geometry}

\usepackage{microtype}
\usepackage{lineno}  
\usepackage{xspace} 
\usepackage{caption} 

\usepackage{graphicx}  
\usepackage{color}
\usepackage{colortbl}
\graphicspath{{./figs/}} 

\usepackage{amsmath} 
\usepackage{amssymb}
\usepackage{amsfonts}
\usepackage{upgreek} 

\newcommand*\patchAmsMathEnvironmentForLineno[1]{%
\expandafter\let\csname old#1\expandafter\endcsname\csname #1\endcsname
\expandafter\let\csname oldend#1\expandafter\endcsname\csname
end#1\endcsname
 \renewenvironment{#1}%
   {\linenomath\csname old#1\endcsname}%
   {\csname oldend#1\endcsname\endlinenomath}%
}
\newcommand*\patchBothAmsMathEnvironmentsForLineno[1]{%
  \patchAmsMathEnvironmentForLineno{#1}%
  \patchAmsMathEnvironmentForLineno{#1*}%
}
\AtBeginDocument{%
\patchBothAmsMathEnvironmentsForLineno{equation}%
\patchBothAmsMathEnvironmentsForLineno{align}%
\patchBothAmsMathEnvironmentsForLineno{flalign}%
\patchBothAmsMathEnvironmentsForLineno{alignat}%
\patchBothAmsMathEnvironmentsForLineno{gather}%
\patchBothAmsMathEnvironmentsForLineno{multline}%
\patchBothAmsMathEnvironmentsForLineno{eqnarray}%
}

\usepackage{hyperxmp}

\usepackage[pdftex,
            pdfauthor={\paperauthors},
            pdftitle={\paperasciititle},
            pdfkeywords={\paperkeywords},
            pdfcopyright={Copyright (C) \papercopyright},
            pdflicenseurl={\paperlicenceurl}]{hyperref}

\usepackage[bottom,flushmargin,hang,multiple]{footmisc}

\usepackage[all]{hypcap} 

\usepackage{xspace} 
\usepackage{upgreek}

\def\lhcb   {\mbox{LHCb}\xspace}

\def\MagUp {\mbox{\em Mag\kern -0.05em Up}\xspace}

\ifthenelse{\boolean{uprightparticles}}%
{
 
 \def\Pgamma      {\ensuremath{\upgamma}\xspace}

 \def\Ppi         {\ensuremath{\uppi}\xspace}

 \def\Ppsi        {\ensuremath{\uppsi}\xspace}

 \def\PDelta      {\ensuremath{\Delta}\xspace}                 
 \def\PXi         {\ensuremath{\Xi}\xspace}                 
 \def\PLambda     {\ensuremath{\Lambda}\xspace}                 
 \def\PSigma      {\ensuremath{\Sigma}\xspace}                 
 \def\POmega      {\ensuremath{\Omega}\xspace}                 
 \def\PUpsilon    {\ensuremath{\Upsilon}\xspace}
 \let\oldPi\Pi
 \def\PPi         {\ensuremath{\oldPi}\xspace}

 \def\PB      {\ensuremath{\mathrm{B}}\xspace}                 
                  
 \def\PD      {\ensuremath{\mathrm{D}}\xspace}

 \def\PJ      {\ensuremath{\mathrm{J}}\xspace}                 
 \def\PK      {\ensuremath{\mathrm{K}}\xspace}

 \def\PP      {\ensuremath{\mathrm{P}}\xspace}

 \def\PW      {\ensuremath{\mathrm{W}}\xspace}

 \def\Pb      {\ensuremath{\mathrm{b}}\xspace}                 
 \def\Pc      {\ensuremath{\mathrm{c}}\xspace}                 
 \def\Pd      {\ensuremath{\mathrm{d}}\xspace}

 \def\Pi      {\ensuremath{\mathrm{i}}\xspace}

 \def\Pp      {\ensuremath{\mathrm{p}}\xspace}

 \def\Ps      {\ensuremath{\mathrm{s}}\xspace}                 
                  
 \def\Pu      {\ensuremath{\mathrm{u}}\xspace}

 \def\thebaroffset{0.0em}
}
{
 
 \def\Pgamma      {\ensuremath{\gamma}\xspace}

 \def\Ppi         {\ensuremath{\pi}\xspace}

 \def\Ppsi        {\ensuremath{\psi}\xspace}                 
                  
 \mathchardef\PDelta="7101
 \mathchardef\PXi="7104
 \mathchardef\PLambda="7103
 \mathchardef\PSigma="7106
 \mathchardef\POmega="710A
 \mathchardef\PUpsilon="7107
 \mathchardef\PPi="7105
                  
 \def\PB      {\ensuremath{B}\xspace}                 
                  
 \def\PD      {\ensuremath{D}\xspace}

 \def\PJ      {\ensuremath{J}\xspace}                 
 \def\PK      {\ensuremath{K}\xspace}

 \def\PP      {\ensuremath{P}\xspace}

 \def\PW      {\ensuremath{W}\xspace}

 \def\Pb      {\ensuremath{b}\xspace}                 
 \def\Pc      {\ensuremath{c}\xspace}                 
 \def\Pd      {\ensuremath{d}\xspace}

 \def\Pi      {\ensuremath{i}\xspace}

 \def\Pp      {\ensuremath{p}\xspace}

 \def\Ps      {\ensuremath{s}\xspace}                 
                  
 \def\Pu      {\ensuremath{u}\xspace}

 \def\thebaroffset{0.18em}
}
\newcommand{\offsetoverline}[2][\thebaroffset]{\kern #1\overline{\kern -#1 #2}}%

\makeatletter
\ifcase \@ptsize \relax
  \newcommand{\miniscule}{\@setfontsize\miniscule{4}{5}}
\or
  \newcommand{\miniscule}{\@setfontsize\miniscule{5}{6}}
\or
  \newcommand{\miniscule}{\@setfontsize\miniscule{5}{6}}
\fi
\makeatother

\DeclareRobustCommand{\optbar}[1]{\shortstack{{\miniscule (\rule[.5ex]{1.25em}{.18mm})}
  \\ [-.7ex] $#1$}}

\def\g      {{\ensuremath{\Pgamma}}\xspace}

\def\W      {{\ensuremath{\PW}}\xspace}

\def\Wm     {{\ensuremath{\PW^-}}\xspace}

\def\uquark    {{\ensuremath{\Pu}}\xspace}
\def\uquarkbar {{\ensuremath{\overline \uquark}}\xspace}

\def\dquark    {{\ensuremath{\Pd}}\xspace}

\def\squark    {{\ensuremath{\Ps}}\xspace}

\def\cquark    {{\ensuremath{\Pc}}\xspace}
\def\cquarkbar {{\ensuremath{\overline \cquark}}\xspace}

\def\bquark    {{\ensuremath{\Pb}}\xspace}

\def\pion   {{\ensuremath{\Ppi}}\xspace}
\def\piz    {{\ensuremath{\pion^0}}\xspace}
\def\pip    {{\ensuremath{\pion^+}}\xspace}
\def\pim    {{\ensuremath{\pion^-}}\xspace}

\def\kaon    {{\ensuremath{\PK}}\xspace}

\def\KorKbar {\kern \thebaroffset\optbar{\kern -\thebaroffset \PK}{}\xspace}

\def\Kp      {{\ensuremath{\kaon^+}}\xspace}
\def\Km      {{\ensuremath{\kaon^-}}\xspace}

\def\Dbar    {{\ensuremath{\offsetoverline{\PD}}}\xspace}
\def\D       {{\ensuremath{\PD}}\xspace}

\def\DorDbar {\kern \thebaroffset\optbar{\kern -\thebaroffset \PD}\xspace}
\def\Dz      {{\ensuremath{\D^0}}\xspace}
\def\Dzb     {{\ensuremath{\Dbar{}^0}}\xspace}
\def\Dp      {{\ensuremath{\D^+}}\xspace}
\def\Dm      {{\ensuremath{\D^-}}\xspace}

\def\DpDm    {\ensuremath{\Dp {\kern -0.16em \Dm}}\xspace}

\def\Dstarzb {{\ensuremath{\Dbar{}^{*0}}}\xspace}

\def\theDstarzb{{\ensuremath{\Dbar^{*}(2007)^{0}}}\xspace}
\def\Dstarp  {{\ensuremath{\D^{*+}}}\xspace}
\def\Dstarm  {{\ensuremath{\D^{*-}}}\xspace}

\def\Ds      {{\ensuremath{\D^+_\squark}}\xspace}
\def\Dsp     {{\ensuremath{\D^+_\squark}}\xspace}
\def\Dsm     {{\ensuremath{\D^-_\squark}}\xspace}

\def\Dssm    {{\ensuremath{\D^{*-}_\squark}}\xspace}

\def\B       {{\ensuremath{\PB}}\xspace}
\def\Bbar    {{\ensuremath{\offsetoverline{\PB}}}\xspace}

\def\BorBbar {\kern \thebaroffset\optbar{\kern -\thebaroffset \PB}\xspace}

\def\Bzb     {{\ensuremath{\Bbar{}^0}}\xspace}
\def\Bd      {{\ensuremath{\B^0}}\xspace}

\def\BdorBdbar {\kern \thebaroffset\optbar{\kern -\thebaroffset \Bd}\xspace}

\def\Bub     {{\ensuremath{\B^-}}\xspace}

\def\Bm      {{\ensuremath{\Bub}}\xspace}

\def\Bs      {{\ensuremath{\B^0_\squark}}\xspace}

\def\BsorBsbar {\kern \thebaroffset\optbar{\kern -\thebaroffset \Bs}\xspace}

\def\Bcm     {{\ensuremath{\B_\cquark^-}}\xspace}

\def\jpsi     {{\ensuremath{{\PJ\mskip -3mu/\mskip -2mu\Ppsi}}}\xspace}

\def\Y#1S{\ensuremath{\PUpsilon{(#1S)}}\xspace}

\def\proton      {{\ensuremath{\Pp}}\xspace}
\def\antiproton  {{\ensuremath{\overline \proton}}\xspace}

\def\Lz          {{\ensuremath{\PLambda}}\xspace}
\def\Lbar        {{\ensuremath{\offsetoverline{\PLambda}}}\xspace}
\def\LorLbar     {\kern \thebaroffset\optbar{\kern -\thebaroffset \PLambda}\xspace}

\def\Sigmares    {{\ensuremath{\PSigma}}\xspace}

\def\Xires       {{\ensuremath{\PXi}}\xspace}

\def\Lc          {{\ensuremath{\Lz^+_\cquark}}\xspace}
\def\Lcbar       {{\ensuremath{\Lbar{}^-_\cquark}}\xspace}

\def\Lb           {{\ensuremath{\Lz^0_\bquark}}\xspace}

\newcommand{\decay}[2]{\ensuremath{#1\!\to #2}\xspace} 

\def\to                 {\ensuremath{\rightarrow}\xspace}

\def\AT#1     {\ensuremath{A_{\mathrm{T}}^{#1}}\xspace}           

\def\C#1      {\ensuremath{\mathcal{C}_{#1}}\xspace}                       
\def\Cp#1     {\ensuremath{\mathcal{C}_{#1}^{'}}\xspace}                    
\def\Ceff#1   {\ensuremath{\mathcal{C}_{#1}^{\mathrm{(eff)}}}\xspace}        
\def\Cpeff#1  {\ensuremath{\mathcal{C}_{#1}^{'\mathrm{(eff)}}}\xspace}       
\def\Ope#1    {\ensuremath{\mathcal{O}_{#1}}\xspace}                       
\def\Opep#1   {\ensuremath{\mathcal{O}_{#1}^{'}}\xspace}                    

\newcommand{\nospaceunit}[1]{\ensuremath{\text{#1}}}       
\newcommand{\aunit}[1]{\ensuremath{\text{\,#1}}}       

\newcommand{\tev}{\aunit{Te\kern -0.1em V}\xspace}
\newcommand{\gev}{\aunit{Ge\kern -0.1em V}\xspace}
\newcommand{\mev}{\aunit{Me\kern -0.1em V}\xspace}
\newcommand{\kev}{\aunit{ke\kern -0.1em V}\xspace}
\newcommand{\ev}{\aunit{e\kern -0.1em V}\xspace}
 
\newcommand{\mevc}{\ensuremath{\aunit{Me\kern -0.1em V\!/}c}\xspace}
\newcommand{\gevc}{\ensuremath{\aunit{Ge\kern -0.1em V\!/}c}\xspace}
\newcommand{\mevcc}{\ensuremath{\aunit{Me\kern -0.1em V\!/}c^2}\xspace}
\newcommand{\gevcc}{\ensuremath{\aunit{Ge\kern -0.1em V\!/}c^2}\xspace}

\def\mum  {\ensuremath{\,\upmu\nospaceunit{m}}\xspace}

\def\fb   {\ensuremath{\aunit{fb}}\xspace}
\def\invfb   {\ensuremath{\fb^{-1}}\xspace}

\newcommand{\chisqip}{\ensuremath{\chi^2_{\text{IP}}}\xspace}

\def\gsim{{~\raise.15em\hbox{$>$}\kern-.85em
          \lower.35em\hbox{$\sim$}~}\xspace}
\def\lsim{{~\raise.15em\hbox{$<$}\kern-.85em
          \lower.35em\hbox{$\sim$}~}\xspace}

\def\sqs   {\ensuremath{\protect\sqrt{s}}\xspace}

\def\pt         {\ensuremath{p_{\mathrm{T}}}\xspace}

\def\ptot       {\ensuremath{p}\xspace}

\def\evtgen     {\mbox{\textsc{EvtGen}}\xspace}

\def\geant      {\mbox{\textsc{Geant4}}\xspace}

\def\photos     {\mbox{\textsc{Photos}}\xspace}

\def\pythia     {\mbox{\textsc{Pythia}}\xspace}

\def\tell1  {TELL1\xspace}
\def\ukl1   {UKL1\xspace}

\newcommand{\eg}{\mbox{\itshape e.g.}\xspace}
\newcommand{\ie}{\mbox{\itshape i.e.}\xspace}

\newcommand{\lhcborcid}[1]{\href{https://orcid.org/#1}{\hspace*{0.1em}\raisebox{-0.45ex}{\includegraphics[width=1em]{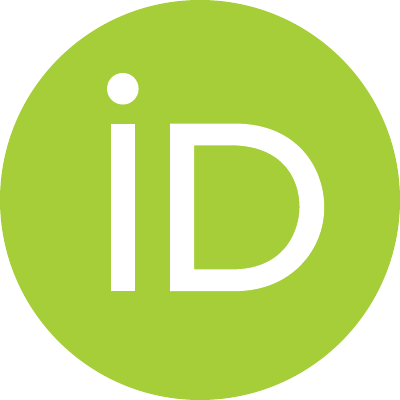}}}}

\usepackage{cite} 
\usepackage{mciteplus}

\usepackage{tabularx,booktabs,adjustbox,multirow} 
\usepackage{tikz}
\usepackage[compat=1.1.0]{tikz-feynman}
\tikzfeynmanset{warn luatex=false}
\tikzstyle{every picture}+=[remember picture] \everymath{\displaystyle}
\usetikzlibrary{external,arrows,shapes,positioning,patterns,plotmarks}%
\usepackage[hypcap=true,labelformat=simple]{subcaption}

\graphicspath{{figs}} 

\makeatletter
\def\input@path{{tables},{figs}}
\makeatother

\makeatletter
\newcommand{\xRightarrow}[2][]{\ext@arrow 0359\Rightarrowfill@{#1}{#2}}
\makeatother

\newcommand{\BFRatio}[2]{\frac{\mathcal{B}\left(#1\right)}{\mathcal{B}\left(#2\right)}}
\newcommand{\BFRatioinline}[2]{\mathcal{B}\left(#1\right)/\mathcal{B}\left(#2\right)}

\providecommand{\DzborDstarzb}{\Dbar{}^{(*)0}}

\providecommand{\DsmorDssm}{\D^{(*)-}_\squark}

\providecommand{\Dsttstm}{\D^*_{\squark 0}(2317)^-}

\providecommand{\Dstfsm}{\D_{\squark 1}(2460)^-}
\providecommand{\Dstftsm}{\D_{\squark 1}(2536)^-}
\providecommand{\Dstshm}{\D^*_{\squark 1}(2700)^-}

\providecommand{\Lcs}{\Lz_\cquark^{*+}}
\providecommand{\Lctfnfp}{\Lz_\cquark(2595)^+}
\providecommand{\Lctstfp}{\Lz_\cquark(2625)^+}

\providecommand{\Scp}{\Sigmares_\cquark^+}

\providecommand{\Sctfffp}{\Sigmares_\cquark(2455)^+}

\providecommand{\Sctftp}{\Sigmares_\cquark(2520)^+}

\providecommand{\Xictsnz}{\Xires_\cquark(2790)^0}

\providecommand{\Pcp}{\PP_\cquark^+}

\providecommand{\Kppim}{\Kp\pim}

\providecommand{\pKpi}{\proton\Km\pip}
\providecommand{\KKpim}{\Km\Kp\pim}
\providecommand{\LcpKpi}{\decay{\Lc}{\pKpi}}

\providecommand{\DzbKpi}{\decay{\Dzb}{\Kp\pim}}
\providecommand{\DsKKpi}{\decay{\Dsp}{\Kp\Km\pip}}
\providecommand{\DsmKKpi}{\decay{\Dsm}{\KKpim}}

\providecommand{\LbLcDzbK}{\decay{\Lb}{\Lc\Dzb\Km}}
\providecommand{\LbLcDs}{\decay{\Lb}{\Lc\Dsm}}
\providecommand{\LcDzbK}{\Lc\Dzb\Km}
\providecommand{\LcDs}{\Lc\Dsm}

\providecommand{\LcDstzb}{\Lc\Dstarzb}

\providecommand{\LcDstzbK}{\LcDstzb\Km}
\providecommand{\LbLcDstzbK}{\decay{\Lb}{\LcDstzbK}}
\providecommand{\LbLcDzbstK}{\decay{\Lb}{\Lc\DzborDstarzb\Km}}
\providecommand{\LbLcDsorsm}{\decay{\Lb}{\Lc\DsmorDssm}}
\providecommand{\LbScDzbK}{\decay{\Lb}{\Scp\Dzb\Km}}
\providecommand{\LbLcDss}{\decay{\Lb}{\Lc\Dssm}}

\providecommand{\LbLcDssz}{\decay{\Lb}{\Lc\Dsttstm}}
\providecommand{\LbLcDsso}{\decay{\Lb}{\Lc\Dstfsm}}
\providecommand{\LbLcDspipi}{\decay{\Lb}{\Lc\Dsm\pi\pi}}

\providecommand{\LbLcDssPRg}{\decay{\Lb}{\Lc\left[\Dsm\g\right]_{\Dssm}}}
\providecommand{\LbLcDssPRpi}{\decay{\Lb}{\Lc\left[\Dsm\piz\right]_{\Dssm}}}

\providecommand{\LbLcDsszPRpi}{\decay{\Lb}{\Lc\left[\Dsm\piz\right]_{\Dsttstm}}}
\providecommand{\LbLcDssoPRg}{\decay{\Lb}{\Lc\left[\Dsm\g\right]_{\Dstfsm}}}
\providecommand{\LbLcsDsPRpipi}{\decay{\Lb}{\left[\Lc\pi\pi\right]_{\Lcs}\Dsm}}
\providecommand{\LbLcDstzbPRpiz}{\decay{\Lb}{\Lc\left[\Dzb\piz\right]_{\Dstarzb}\Km}}
\providecommand{\LbLcDstzbPRg}{\decay{\Lb}{\Lc\left[\Dzb\g\right]_{\Dstarzb}\Km}}

\providecommand{\LbSctfffpDzbPRpiz}{\decay{\Lb}{\left[\Lc\piz\right]_{\Sctfffp}\Dzb\Km}}

\providecommand{\LbLcKpiK}{\decay{\Lb}{\Lc\Kppim\Km}}

\providecommand{\LbLcpi}{\decay{\Lb}{\Lc\pim}}

\providecommand{\LbJpK}{\decay{\Lb}{\jpsi\proton\Km}}

\begin{document}

\renewcommand{\thefootnote}{\fnsymbol{footnote}}
\setcounter{footnote}{1}


\begin{titlepage}
\pagenumbering{roman}

\vspace*{-1.5cm}
\centerline{\large EUROPEAN ORGANIZATION FOR NUCLEAR RESEARCH (CERN)}
\vspace*{1.5cm}
\noindent
\begin{tabular*}{\linewidth}{lc@{\extracolsep{\fill}}r@{\extracolsep{0pt}}}
\ifthenelse{\boolean{pdflatex}}
{\vspace*{-1.5cm}\mbox{\!\!\!\includegraphics[width=.14\textwidth]{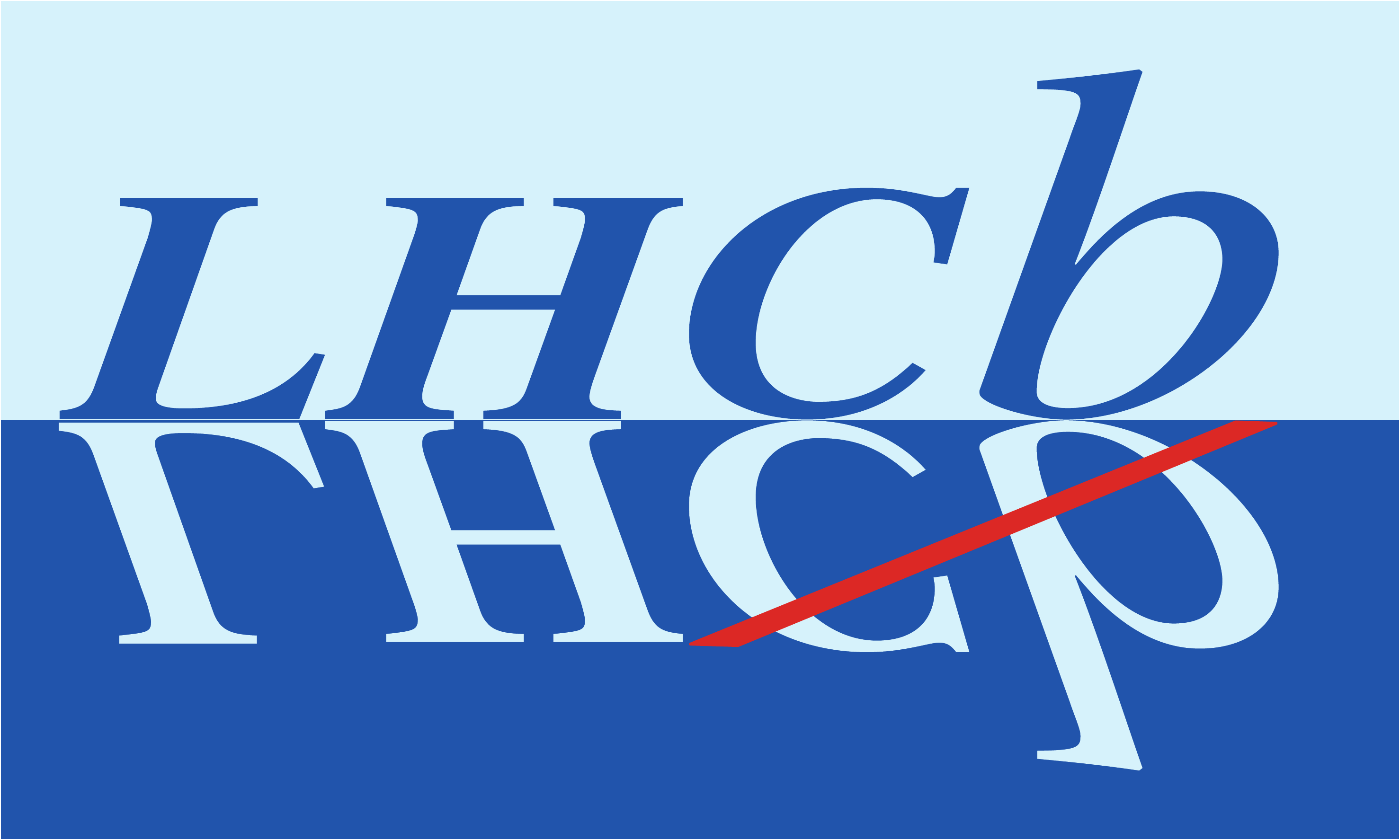}} & &}%
{\vspace*{-1.2cm}\mbox{\!\!\!\includegraphics[width=.12\textwidth]{lhcb-logo.eps}} & &}%
\\
 & & CERN-EP-2023-244 \\  
 & & LHCb-PAPER-2023-034 \\  
 & & June 06, 2024 \\ 
 & & \\
\end{tabular*}

\vspace*{3.0cm}

{\normalfont\bfseries\boldmath\huge
\begin{center}
  \papertitle
\end{center}
}

\vspace*{0.4cm}

\begin{center}
\paperauthors\footnote{Authors are listed at the end of this paper.}
\end{center}

\vspace{\fill}

\begin{abstract}
\vspace*{0.2cm}
  \noindent
  The decays $\LbLcDzbstK$ and $\LbLcDss$ are observed for the first time, in proton-proton collision data at $\sqs=13\tev$, corresponding to an integrated luminosity of 5.4\invfb collected with the \lhcb detector.
  Their ratios of branching fractions with respect to the $\LbLcDs$ mode are measured to be
  \begin{align*}
    \begin{split}
      \BFRatio{\LbLcDzbK}{\LbLcDs}   & = 0.1908 {}_{-0.0034}^{+0.0036} {}_{-0.0018}^{+0.0016} \pm 0.0038,\\
      \BFRatio{\LbLcDstzbK}{\LbLcDs} & = 0.589 {}_{-0.017}^{+0.018} {}_{-0.018}^{+0.017} \pm 0.012,\\
      \BFRatio{\LbLcDss}{\LbLcDs}    & = 1.668 \pm 0.022 {}_{-0.055}^{+0.061},
    \end{split}
  \end{align*}
  where the first uncertainties are statistical, the second systematic, and the third, for the $\LbLcDzbstK$ decays, are due to the uncertainties on the branching fractions of the $\DsmKKpi$ and $\DzbKpi$ decay modes. The measured branching fractions probe factorization assumptions in effective theories and provide the normalization for future pentaquark searches in $\LbLcDzbstK$ decay channels.
\end{abstract}

\vspace*{1.0cm}

\begin{center}
  Published in EPJC 84 (2024) 575
\end{center}

\vspace{\fill}

{\footnotesize
\centerline{\copyright~\papercopyright. \href{\paperlicenceurl}{\paperlicence}.}}
\vspace*{2mm}

\end{titlepage}


\newpage
\setcounter{page}{2}
\mbox{~}
%


\renewcommand{\thefootnote}{\arabic{footnote}}
\setcounter{footnote}{0}


\pagestyle{plain} 
\setcounter{page}{1}
\pagenumbering{arabic}


\section{Introduction}\label{sec:Introduction}
Hadrons are systems of quarks bound by the strong interaction, described at the
fundamental level by quantum chromodynamics (QCD).
Decays of heavy hadrons containing at least one $b$ quark provide clean signatures to test new emergent phenomena
in the realms of QCD and physics beyond the Standard Model.
Effective theories testing these signatures are based on the fact that the heavy-quark mass $m_Q$ (\eg $m_\bquark\sim 4\gev$) is much
larger than the QCD scale $\Lambda_\text{QCD}\sim 100\mev$.\footnote{The inclusion of charge conjugate processes and the use of natural units are implicit throughout this article.} Processes that occur at the scale of $m_Q$ can be described by perturbative QCD, while nonperturbative effects, including the
formation of light hadrons, are suppressed by powers of \mbox{$\Lambda_\text{QCD}/m_Q$}.
This factorization of high-energy and low-energy processes is widely used in effective theories describing decays of heavy hadrons, such as
the heavy quark effective theory (HQET)~\cite{Isgur:1989vq,Isgur:1990yhj,Eichten:1989zv,Neubert:1993mb}.

Beauty hadron decays to two charmed hadrons probe factorization assumptions in HQET in a regime where
their application is contestable due to the presence of two charm quarks (\mbox{$m_\cquark\sim1.3\gev$}) in the final state.
In particular, decays of the $\Lb$ baryon, which has quark content $udb$, are interesting as the $(\uquark\dquark)$ diquark is in a spin-zero state, which gives rise to additional symmetries of HQET. In that sense, the $\Lb$ baryon is a simpler object than a $B$ meson.

This article reports the measurement of the branching fraction of \mbox{$\LbLcDzbK$},
\mbox{$\LbLcDstzbK$} and \mbox{$\LbLcDss$} double-open-charm decays, relative to that of the \mbox{$\LbLcDs$} decay. The dominant Feynman diagrams of these decays are shown in Fig~\ref{fig:feynman}, where the left and middle diagrams contribute to \mbox{$\LbLcDzbstK$} decays, while the right diagram corresponds to \mbox{$\LbLcDsorsm$} decays.
The color-suppressed, internal $\W$-emission diagram, as shown in the middle, does not exist for \mbox{$\LbLcDsorsm$} decays. The isospin of the light diquark in the left and right diagrams is conserved, such that \mbox{$\decay{\Lb}{\Scp\DzborDstarzb\Km}$} decays are suppressed, and \mbox{$\decay{\Lb}{\Scp\DsmorDssm}$} decays are forbidden up to nonfactorizing contributions~\cite{Datta:1998ex, Datta:2003yk}.

\begin{figure}[!h]
  \begin{adjustbox}{width=0.325\textwidth}
    \begin{tikzpicture}
  \begin{feynman}
    \vertex                   (b1) {\(\bquark\)};
    \vertex[below= 2em of b1] (d1) {\(\dquark\)};
    \vertex[below= 2em of d1] (u1) {\(\uquark\)};
    \vertex[right= 2cm of b1] (W1);
    \vertex[right= 5cm of u1] (u2) {\(\uquark\)};
    \vertex[right= 5cm of b1] (c1) {\(\cquark\)};
    \vertex[right= 5cm of d1] (d2) {\(\dquark\)};
    \vertex[above= 2em of c1] (c2) {\(\cquarkbar\)};
    \vertex[above= 6em of c2] (s1) {\(\squark\)};
    \vertex[above= 2em of c2] (u3) {\(\uquark\)};
    \vertex[above= 2em of u3] (u4) {\(\uquarkbar\)};
    \vertex at ($(s1)!0.5!(c2) - (1.5cm, 0)$) (W2);
    \diagram* {
      {[edges=fermion]
        (b1) -- (W1) -- (c1),
        (u1) -- (u2),
        (d1) -- (d2)
      },
      (s1) -- [fermion, out=180, in=89] (W2) -- [fermion, out=-89, in=180] (c2),
      (W1) -- [boson, bend left,edge label=\(\Wm\)] (W2),
      (u4) -- [out=180, in=180]  (u3),
    };
    \draw [decoration={brace}, decorate] (u1.south west) -- (b1.north west)
          node [pos=0.5, left] {\(\Lb\)};
    \draw [decoration={brace}, decorate] (s1.north east) -- (u4.south east)
           node [pos=0.5, right] {\(\Km\)};
    \draw [decoration={brace}, decorate] (u3.north east) -- (c2.south east)
          node [pos=0.5, right] {\(\DzborDstarzb\)};
    \draw [decoration={brace}, decorate] (c1.north east) -- (u2.south east)
          node [pos=0.5, right] {\(\Lc\)};
  \end{feynman}
\end{tikzpicture}
  \end{adjustbox}
  \begin{adjustbox}{width=0.325\textwidth}
    \begin{tikzpicture}
  \begin{feynman}
    \vertex                   (u1) {\(\uquark\)};
    \vertex[below=10em of u1] (b1) {\(\bquark\)};
    \vertex[below= 2em of b1] (d1) {\(\dquark\)};
    \vertex[right= 2cm of b1] (W1);
    \vertex[right= 5cm of u1] (u2) {\(\uquark\)};
    \vertex[right= 5cm of b1] (c1) {\(\cquark\)};
    \vertex[right= 5cm of d1] (d2) {\(\dquark\)};
    \vertex[below= 2em of u2] (c2) {\(\cquarkbar\)};
    \vertex[below= 2em of c2] (s1) {\(\squark\)};
    \vertex[below= 2em of s1] (u3) {\(\uquarkbar\)};
    \vertex[below= 2em of u3] (u4) {\(\uquark\)};
    \vertex at ($(s1)!0.5!(c2) - (1.5cm, 0)$) (W2);
    \diagram* {
      {[edges=fermion]
        (b1) -- (W1) -- (c1),
        (u1) -- (u2),
        (d1) -- (d2)
      },
      (c2) -- [fermion, out=180, in=85] (W2) -- [fermion, out=-85, in=180] (s1),
      (W1) -- [boson, bend left,edge label=\(\Wm\)] (W2),
      (u4) -- [out=180, in=180]  (u3),
    };
    \draw [decoration={brace}, decorate] (d1.south west) -- (u1.north west)
          node [pos=0.5, left] {\(\Lb\)};
    \draw [decoration={brace}, decorate] (s1.north east) -- (u3.south east)
           node [pos=0.5, right] {\(\Km\)};
    \draw [decoration={brace}, decorate] (u2.north east) -- (c2.south east)
          node [pos=0.5, right] {\(\DzborDstarzb\)};
    \draw [decoration={brace}, decorate] (u4.north east) -- (d2.south east)
          node [pos=0.5, right] {\(\Lc\) or \(\Scp\)};
  \end{feynman}
\end{tikzpicture}
  \end{adjustbox}
  \begin{adjustbox}{width=0.325\textwidth}
    \begin{tikzpicture}
  \begin{feynman}
    \vertex                   (b1) {\(\bquark\)};
    \vertex[below= 2em of b1] (d1) {\(\dquark\)};
    \vertex[below= 2em of d1] (u1) {\(\uquark\)};
    \vertex[right= 2cm of b1] (W1);
    \vertex[right= 5cm of u1] (u2) {\(\uquark\)};
    \vertex[right= 5cm of b1] (c1) {\(\cquark\)};
    \vertex[right= 5cm of d1] (d2) {\(\dquark\)};
    \vertex[above= 2em of c1] (s1) {\(\squark\)};
    \vertex[above= 2em of s1] (c2) {\(\cquarkbar\)};    
    \vertex at ($(s1)!0.5!(c2) - (1.5cm, 0)$) (W2);
    \diagram* {
      {[edges=fermion]
        (b1) -- (W1) -- (c1),
        (u1) -- (u2),
        (d1) -- (d2)
      },
      (c2) -- [fermion, out=180, in=85] (W2) -- [fermion, out=-85, in=180] (s1),
      (W1) -- [boson, bend left,edge label=\(\Wm\)] (W2),      
    };
    \draw [decoration={brace}, decorate] (u1.south west) -- (b1.north west)
          node [pos=0.5, left] {\(\Lb\)};
    \draw [decoration={brace}, decorate] (c2.north east) -- (s1.south east)
           node [pos=0.5, right] {\(\DsmorDssm\)};
    \draw [decoration={brace}, decorate] (c1.north east) -- (u2.south east)
          node [pos=0.5, right] {\(\Lc\)};
  \end{feynman}
\end{tikzpicture}
  \end{adjustbox}
  \caption[]{Dominant Feynman diagrams for (left, middle) the \mbox{$\LbLcDzbstK$} decays and (right) the \mbox{$\LbLcDsorsm$} decays.}\label{fig:feynman}
\end{figure}

Two-body beauty to double-open-charm decays have been theoretically studied for over three decades, and several
models predict the ratio of branching fractions \mbox{$\BFRatioinline{\LbLcDss}{\LbLcDs}$} in the range 0.75--2.2~\cite{Mannel:1992ti,Cheng:1996cs,Giri:1997te,Fayyazuddin:1998ap,Mohanta:1998iu,Zhu:2018jet,Zhao:2018zcb,Liang:2018rkl,Gutsche:2018utw,Ke:2019smy,Chua:2019yqh,Rahmani:2020kjd,Pan:2023hrk}.
Due to the complexity of the three-body system, there are no predictions for the \mbox{$\LbLcDzbstK$} channels.
However, the contribution of color-suppressed amplitudes in \mbox{$\LbLcDzbstK$} can be qualitatively assessed by comparing the relative branching fractions to those of mesons, when exchanging the $ud$ diquark in the $\Lb$ with an antiquark~\cite{Burns:2022uiv}. For this comparison, it is convenient to define the double ratio
\begin{equation}
  \mathcal{DR}^{(*)}(M_b) \equiv \left[ \BFRatio{\LbLcDzbstK}{\Lb\to\Lc\Dsm}\right]\Big/\left[ \BFRatio{M_b\to M_c \DzborDstarzb\Km}{M_b\to M_c \Dsm}\right],
\label{eq:dr}
\end{equation}
where $M_b$ or $M_c$ is a beauty or charm meson, and the star in $\mathcal{DR}^{(*)}$ denotes the $\Dzb$ ground state or the $\theDstarzb$ vector state.
Additionally, there is a large interest in exploring the resonance structure of the \mbox{$\LbLcDzbstK$} decays. The reason is that the $\Lc\Dzb$ and $\Lc\Dstarzb$ systems are the open-charm
equivalent of  the $\jpsi(\cquark\cquarkbar)\proton(\uquark\uquark\dquark)$ final state,
where $\Pcp(\cquark\cquarkbar\uquark\uquark\dquark)$ pentaquark resonances have
been observed~\cite{LHCb-PAPER-2015-029, LHCb-PAPER-2016-009, LHCb-PAPER-2016-015, LHCb-PAPER-2019-014}.
Decays of these pentaquarks to $\Lc\Dzb$ and $\Lc\Dstarzb$ states are anticipated in many models, yet the predicted branching fractions relative to that into $\jpsi\proton$ vary by orders of magnitude~\cite{Guo:2019kdc,Xiao:2019aya,Weng:2019ynv,Voloshin:2019aut,Lin:2019qiv,Semenova:2019gzf,Chen:2020pac,Dong:2020nwk,Wang:2020rdh,Xiao:2020frg,Li:2023aui,Pan:2023hrk}.
To test those predictions experimentally, two more ingredients are needed, namely the pentaquark fit fractions coming from \mbox{$\LbLcDstzbK$}
amplitude analyses $f_{\Lc\DzborDstarzb}(\Pcp)$, and the branching fractions of \mbox{$\LbLcDzbstK$} decays relative to that of the \mbox{$\LbJpK$} mode~\cite{LHCb-PAPER-2015-032}.
Assuming that the production mechanism for \mbox{$\decay{\Lb}{\Pcp\Km}$} is the same as for \mbox{$\LbLcDzbstK$} and \mbox{$\LbJpK$} decays, the fit fraction in the \mbox{$\LbLcDzbstK$} final states is given by
\begin{equation}
  f_{\Lc\DzborDstarzb}(\Pcp) = f_{\jpsi\proton}(\Pcp) \cdot \BFRatio{\decay{\Lb}{\jpsi\proton\Km}}{\LbLcDzbstK} \cdot \BFRatio{\decay{\Pcp}{\Lc\DzborDstarzb}}{\decay{\Pcp}{\jpsi\proton}}.
  \label{eq:pcbf}
\end{equation}
Thus, the values of $\BFRatio{\decay{\Lb}{\jpsi\proton\Km}}{\LbLcDzbstK}$ that will be derived in this article, can be used to calculate sensitivities for observing $\Pcp$ in the $\Lc\DzborDstarzb$ system for a given theoretical prediction of $\BFRatio{\decay{\Pcp}{\Lc\DzborDstarzb}}{\decay{\Pcp}{\jpsi\proton}}$.

Using 5.4\invfb of proton-proton ($pp$) collision data collected at $\sqs=13\tev$ by the LHCb detector in 2015--2018, the branching fractions of \mbox{$\LbLcDzbK$}, \mbox{$\LbLcDstzbK$}, and \mbox{$\LbLcDss$} decays are measured relative to that of $\LbLcDs$ decays.
The latter is chosen as a normalization channel because it has large signal yield, and its branching fraction has been previously measured~\cite{LHCb-PAPER-2014-002}.
The $\Lb$ decays proceed through different intermediate charm hadron decays, namely \mbox{$\LcpKpi$}, $\DzbKpi$ and/or $\DsmKKpi$.
All decays are reconstructed with the same particles in the final state, $\proton\Km\Km\Kp\pim\pip$, which reduces or cancels various uncertainties related to the $\Lb$ production and the determination of absolute reconstruction and selection efficiencies.
Neutral objects or electron-positron pairs, in the decays of $\Dstarzb$ or $\Dssm$ mesons are not reconstructed. Therefore, the \mbox{$\LbLcDstzbK$} and \mbox{$\LbLcDss$} decays are referred to as partially reconstructed signal.

The expressions for the ratio of branching fractions for \mbox{$\LbLcDzbstK$} and \mbox{$\LbLcDss$} decays are
\begin{align*}
  \BFRatio{\LbLcDzbstK}{\LbLcDs} &= \frac{N^{\LbLcDzbstK}}{N^{\LbLcDs}} \frac{\epsilon^{\LbLcDs}}{\epsilon^{\LbLcDzbstK}} \BFRatio{\DsmKKpi}{\DzbKpi}, \\
  \BFRatio{\LbLcDss}{\LbLcDs} &= \frac{N^{\LbLcDss}}{N^{\LbLcDs}} \frac{\epsilon^{\LbLcDs}}{\epsilon^{\LbLcDss}},  
\end{align*}
where $N^X$ are the observed yields and $\epsilon^X$ the efficiency for the decay mode $X$.
The ratio of branching fractions \mbox{$\BFRatioinline{\DsmKKpi}{\DzbKpi}$}
is taken from Ref.~\cite{PDG2022}. 
%

\section{LHCb detector}\label{sec:detector}
The \lhcb detector~\cite{LHCb-DP-2008-001,LHCb-DP-2014-002} is a single-arm forward spectrometer covering the \mbox{pseudorapidity} range \mbox{$2<\eta <5$} at the LHC,
designed for the study of particles containing \bquark or \cquark
quarks. The detector includes a high-precision tracking system
consisting of a silicon-strip vertex detector surrounding the $pp$
interaction region~\cite{LHCb-DP-2014-001}, a large-area silicon-strip detector located
upstream of a dipole magnet with a bending power of about
$4{\mathrm{\,Tm}}$, and three stations of silicon-strip detectors and straw
drift tubes~\cite{LHCb-DP-2013-003,LHCb-DP-2017-001} placed downstream of the magnet.
The tracking system provides a measurement of the momentum, \ptot, of charged particles with
a relative uncertainty that varies from about 0.5\% below 20\gev to 1.0\% at 200\gev.
The minimum distance of a track to a primary vertex (PV), the impact parameter (IP),
is measured with a resolution of \mbox{$(15+29/\pt)\mum$},
where \pt is the component of the momentum transverse to the beam, in\,\gev.
Different types of charged hadrons are distinguished using information
from two ring-imaging Cherenkov (RICH) detectors~\cite{LHCb-DP-2012-003}.
Hadrons are identified by a calorimeter system consisting of
scintillating-pad and preshower detectors, an electromagnetic and a hadronic calorimeter. Muons are identified by a
system composed of alternating layers of iron and multiwire proportional chambers~\cite{LHCb-DP-2012-002}. \\
\indent The online event selection is performed by a trigger~\cite{LHCb-DP-2019-001,LHCb-DP-2019-002},
which consists of a hardware stage, based on information from the calorimeter and muon
systems, followed by a software stage, which applies a full event
reconstruction.\\
\indent Simulation is required to calculate reconstruction and selection efficiencies,
and to determine the shapes of partially reconstructed components in the invariant mass distributions.
Some of these components are modeled by the fast simulation package 
\textsc{RapidSim}~\cite{Cowan:2016tnm} and the \textsc{AmpGen} generator~\cite{AmpGen}.
In the full detector simulation, $pp$ collisions are generated using
\pythia~\cite{Sjostrand:2007gs} with a specific \lhcb
configuration~\cite{LHCb-PROC-2010-056}.  Decays of unstable particles
are described by \evtgen~\cite{Lange:2001uf}, in which final-state
radiation is generated using \photos~\cite{Golonka:2005pn}. The
interaction of the generated particles with the detector, and its response,
are implemented using the \geant
toolkit~\cite{Allison:2006ve, *Agostinelli:2002hh} as described in
Ref.~\cite{LHCb-PROC-2011-006}.
The underlying $pp$ interaction is reused multiple times, with an independently
generated signal decay for each simulated event~\cite{LHCb-DP-2018-004}.

\section{Dataset and selection}\label{sec:selection}
Data are selected first by an online trigger system, consisting of a hardware- and two software stages, further filtered in an offline selection.
Online, data are selected by a sequence of inclusive trigger decisions. The majority of candidates are selected by
criteria based on the topology of $b$-hadron decays~\cite{BBDT,LHCb-PROC-2015-018}.
In the offline selection, candidates that fulfill characteristics of the exclusive \mbox{$\LbLcDzbK$} or $\LbLcDs$ decay are selected using tracks with hits in at least the vertex tracker and the three downstream tracking stations.
Further selection is applied on momentum, transverse momentum, track quality and displacement from any PV.
Intermediate charm hadrons are selected using the distance of closest approach of their decay products, the decay vertex fit $\chi^2$, and the
displacement of the decay vertex from any PV.
Combined information from particle identification (PID) detectors is used to reject topologically similar background contributions.
The $\Lb$ candidates are reconstructed by combining a $\Lc$ candidate with either a $\Dzb$ candidate and a companion kaon or 
a $\Dsm$ candidate.
Kinematic and topological variables are used to suppress combinatorial background: namely the sum of transverse momenta of all the final-state particles,
the presence of at least one track with large displacement from any PV and high momentum and high transverse momentum, the angle between the reconstructed momentum direction of the $\Lb$ candidate and its flight direction and distance, determined from the production and decay vertices.

A clean sample of $\Dzb$ and $\Dsm$ mesons is selected using topological and kinematic criteria, and a requirement on 
the product of the probabilities of final-state kaons and pions to be correctly
identified. These probabilities correspond to the response of a neural network combining PID information from the full detector~\cite{LHCb-DP-2014-002}.
Even though the $\Lc$ decays produce high-momentum protons that can be cleanly reconstructed and identified, the shorter lifetime
compared to $\Dzb$ and $\Dsm$ mesons makes it more difficult to suppress background contributions in the reconstruction of \mbox{$\LcpKpi$} decays.
Consequently, a dedicated gradient boosted decision tree (BDT) algorithm~\cite{Hocker:2007ht} for secondary \mbox{$\LcpKpi$} decays
is trained and calibrated on $\LbLcpi$ data, 
similar to the \mbox{$\LcpKpi$}~BDT classifier used in Ref.~\cite{LHCb-PAPER-2014-002}.
An optimization procedure is carried out, that determines a working point for for the following variables: the \mbox{$\LcpKpi$}~BDT output, the $\chisqip$ of the $\Lb$ candidate, where $\chisqip$ is defined as the difference in the vertex-fit $\chi^2$ of a given PV reconstructed with and without the $\Lb$ candidate, and the probability of the companion kaon to be correctly identified in the \mbox{\Lc\Dzb\Km} channel.
The optimization is carried out in form of a grid search, maximizing the approximate signal significance ($S/\sqrt{S+B}$) multiplied by the purity ($S/(S+B)$). Here, $S$ and $B$ represent the signal and background yields in a 2\,$\sigma$ window around the exclusive $\Lb$ signal mass peak.
This optimization is done independently for signal and normalization channels.

Candidates for which the opening angle between any track pair is smaller than $0.2$~mrad are rejected, which removes artifacts from matching track segments reconstructed with the downstream tracking stations and the vertex locator.
Fiducial regions are selected in the phase-space later used for weighting simulated events, see Sec.~\ref{sec:efficiencies}. This concerns the transverse momentum and pseudorapidity of the $\Lb$ candidate, as well as the total number of tracks in the event.

Sources of peaking background candidates from particle misidentification are studied by statistically subtracting the combinatorial background contribution~\cite{Pivk:2004ty}, identified in a preliminary fit to the invariant mass of the reconstructed system, described in Sec.~\ref{sec:fits}. 
For the background-subtracted sample, invariant-mass distributions with swapped particle hypotheses are investigated to identify and remove candidates with misidentified particles. To increase selection efficiencies, only those candidates that also fail at least one tighter particle identification requirement, are removed as summarized in Table~\ref{tab:vetos}.

\begin{table}[bt]
  \caption[Vetos]{Explicitly rejected physics backgrounds. Some background contributions are present only in the \mbox{$\LcDzbK$} or \mbox{$\LcDs$} systems, while others in both. A particle, $M_\mathrm{misID}$, that decays through a real particle $a$, which is reconstructed as a particle with different mass hypothesis $b$, is denoted as \mbox{$\decay{M_\mathrm{misID}}{\{a \leftarrow b\}X}$}, where $X$ corresponds to the rest of the decay. As there are two $\Km$ mesons in the final state, the subscripts ``$\Lc$'' and ``com'' (for companion) or ``$\Dsm$'' denote the assignment in the \mbox{$\LcDzbK$} or \mbox{$\LcDs$} reconstruction chain. Cases where the proton is misidentified as a pion or kaon and combined into a $\Dz$ candidate, are marked with the $\Dz$ subscript.}
  \begin{footnotesize}
  \begin{tabular}{ccc}
    \toprule
    \mbox{$\LcDzbK$} & \mbox{$\LcDs$} & Both \\\midrule
    $\decay{\phi}{\{\Kp \leftarrow \proton\} K^-_\mathrm{com}}$
    & $\decay{\Dm}{\{\pim \leftarrow K^-_{\Dsm}\} \Kp \pim}$ 
    & $\decay{\phi}{\{\Kp \leftarrow \proton\} K^-_\mathrm{\Lc}}$\\
    $\decay{\Dstarp}{\left[\{\pip \leftarrow \proton\} K^-_\mathrm{com}\right]_{\Dz} \pip}$
    & $\decay{\Lcbar}{\{\antiproton \leftarrow K^-_{\Dsm}\} \Kp \pim}$& $\decay{\D^+_{(\squark)}}{\{\Kp \leftarrow \proton\} K^-_{\Lc} \pip}$ \\
    $\decay{\Dstarp}{\left[\{\Kp \leftarrow \proton\} K^-_\mathrm{com}\right]_{\Dz} \pip}$
    & $\decay{\Lc}{\{\pip \leftarrow \proton\} K^-_{\Lc} \{\proton \leftarrow \pip\}}$
    & $\decay{\Dp}{\{\pip \leftarrow \proton\} K^-_{\Lc} \pip}$\\
    $\decay{\Dstarm}{\{\pim \leftarrow K^-_\mathrm{com}\} \Dzb}$
    & 
    & $\decay{\Dstarp}{\left[\{\pip \leftarrow \proton\} K^-_{\Lc}\right]_{\Dz} \pip}$\\
    $\decay{\Dstarm}{\{\pim \leftarrow K^-_{\Lc}\} \Dzb}$
    & 
    & $\decay{\Dstarp}{\left[\{\Kp \leftarrow \proton\} K^-_{\Lc}\right]_{\Dz} \pip}$\\ \bottomrule   
   \end{tabular}\label{tab:vetos}
   \end{footnotesize}
 \end{table}

Background events from a wrong combination of candidates in the sample of \mbox{$\LcDzbK$} decays, such as \mbox{$\decay{\Lb}{\proton\Dz\Dzb\kaon_{\Lc}^{-}}$},
with \mbox{$\decay{\Dz}{K^-_\mathrm{com}\pip(\piz)}$}, where the subscripts ``$\Lc$'' and ``com'' denote the nominal assignment in the \mbox{$\LcDzbK$} reconstruction chain. are suppressed by
requiring a small IP and $\chisqip$ of the companion kaon $K^-_\mathrm{com}$, with respect to the $\Lb$ decay vertex.

For each species of charm hadron, an invariant-mass requirement is defined such that the central 95\% of the individual charm-candidate signal is retained. These are determined from a three-dimensional fit to the invariant masses of the $\Lb$ candidate and the two intermediate charm-hadron candidates.
To select at most one $\Lb$ candidate for a given LHC bunch crossing, a final selection step randomly removes all but one candidate from each event. This concerns 2.4\% of the \mbox{$\LcDzbK$} candidates and 1.1\% of the \mbox{$\LcDs$} candidates.

\section{Invariant-mass fits}\label{sec:fits}
Unbinned maximum-likelihood fits are first carried out in three mass dimensions, given by the invariant mass of \mbox{$\LcDzbK$} or \mbox{$\LcDs$} decays, and the two charm hadrons. 
The invariant masses of beauty candidates are defined as \mbox{$m(\LcDzbK)\equiv m(\pKpi\Kppim\Km)-m(\pKpi)-m(\Kppim)+M_{\Lc}+M_{\Dzb}$} and \mbox{$m(\LcDs)\equiv m(\pKpi\KKpim)-m(\pKpi)-m(\KKpim)+M_{\Lc}+M_{\Dsm}$}, where $M_X$ is the known value of the mass of particle $X$ from Ref.~\cite{PDG2022}.

This fit is restricted to narrow regions around the exclusively reconstructed \mbox{$\LbLcDzbK$} and \mbox{$\LbLcDs$} signals in any of the three mass dimensions as shown in Fig.~\ref{fig:Fit3D}. The figures illustrate that the three-dimensional fit directly measures the normalization of single-charm and charmless backgrounds. The central interval containing 95\% of the signal component determines the mass selection of the charm candidates, which removes backgrounds of different origin that would complicate model building in a three-dimensional fit in the full mass range. 

\begin{figure}[tb]
  \centering  
  \begin{adjustbox}{width=\textwidth}
    \begin{tikzpicture}
\def\CheckTikzLibraryLoaded#1{ \ifcsname tikz@library@#1@loaded\endcsname \else \PackageWarning{tikz}{usetikzlibrary{#1} is missing in the preamble.} \fi }
\CheckTikzLibraryLoaded{patterns}
\CheckTikzLibraryLoaded{plotmarks}
\definecolor{c}{rgb}{0,0,0};
\draw [anchor= west] (1.5,1.3) node[scale=1.2248, color=c, rotate=0]{Data};
\draw [c,line width=0.9] (0.0,1.3) -- (1.2424,1.3);
\draw [c,line width=0.9] (0.6212,1.34609) -- (0.6212,1.69257);
\draw [c,line width=0.9] (0.6212,1.25786) -- (0.6212,0.91138);
\draw [c,line width=0.9] (0.3416,1.69257) -- (0.9,1.69257);
\draw [c,line width=0.9] (0.3416,0.91138) -- (0.9,0.91138);
\foreach \P in {(0.6212,1.3)}{\draw[mark options={color=c,fill=c},mark size=3.603604pt, line width=1.000000pt, mark=] plot coordinates {\P};}

\draw [anchor= west] (1.5,0) node[scale=1.2248, color=c, rotate=0]{Full model};
\definecolor{c}{RGB}{68,119,170};
\draw [c,line width=2.7] (0.0,0.0) -- (1.3,0.0);

\definecolor{c}{rgb}{0,0,0};
\draw [anchor= west] (6.3,1.3) node[scale=1.2248, color=c, rotate=0]{$\LbLcDs$ or $\LcDzbK$};
\draw [c,dash pattern=on 2.40pt off 2.40pt ,line width=2.7] (4.6,1.3) -- (5.7,1.3);

\draw [anchor= west] (6.3,0.0) node[scale=1.2248, color=c, rotate=0]{$\LbLcKpiK$};
\definecolor{c}{RGB}{204,51,17};
\draw [c, fill=c] (4.6,-0.45) -- (5.9,-0.45) -- (5.9,0.45) -- (4.6,0.45);

\definecolor{c}{rgb}{0,0,0};
\draw [anchor= west] (14.3,1.3) node[scale=1.2248, color=c, rotate=0]{Combinatorial \Dsm or \Dzb};
\definecolor{c}{RGB}{68,187,153};
\draw [c, fill=c] (12.8,0.85) -- (14.1,0.85) -- (14.1,1.75) -- (12.8,1.75);
\definecolor{c}{rgb}{0,0,0};
\draw [anchor= west] (14.3,0.0) node[scale=1.2248, color=c, rotate=0]{Combinatorial \Lc};
\definecolor{c}{RGB}{204,187,68};
\draw [c, fill=c] (12.8,-0.45) -- (14.1,-0.45) -- (14.1,0.45) -- (12.8,0.45);

\definecolor{c}{rgb}{0,0,0};
\draw [anchor= west] (22.0,1.3) node[scale=1.2248, color=c, rotate=0]{Combinatorial background};
\definecolor{c}{rgb}{0.8,0.8,0.8};
\draw [c, fill=c] (20.5,0.85) -- (21.8,0.85) -- (21.8,1.75) -- (20.5,1.75);

\end{tikzpicture}
  \end{adjustbox}\\[2mm]
  \begin{adjustbox}{width=0.49\textwidth}
    \input{fits/LcDsRun2_selA_3D_mLb_for_paper}
  \end{adjustbox}
  \begin{adjustbox}{width=0.49\textwidth}
    \input{fits/LcD0KRun2_selA_3D_mLb_for_paper}
  \end{adjustbox}
  \begin{adjustbox}{width=0.49\textwidth}
    \input{fits/LcDsRun2_selA_3D_Lc_M_for_paper}
  \end{adjustbox}
  \begin{adjustbox}{width=0.49\textwidth}
    \input{fits/LcD0KRun2_selA_3D_Lc_M_for_paper}
  \end{adjustbox}
  \begin{adjustbox}{width=0.49\textwidth}
    \input{fits/LcDsRun2_selA_3D_Ds_M_for_paper}
  \end{adjustbox}
  \begin{adjustbox}{width=0.49\textwidth}
    \input{fits/LcD0KRun2_selA_3D_D0_M_for_paper}
  \end{adjustbox}
  \caption[]{Distributions of (upper left) $m(\LcDs)$, (upper right) $m(\LcDzbK)$, (middle) $m(p\Km\pip)$, (lower left) $m(\Km\Kp\pim)$ and (lower right) $m(\Kp\pim)$ for the (left) $\Lc\Dsm$ and (right) $\LcDzbK$ candidates, with the fit projections overlaid.} \label{fig:Fit3D}
\end{figure}%

Exclusive signal contributions in all mass dimensions are modeled by two-sided Hypatia functions, which are
convolved with a Gaussian function~\cite{Santos:2013gra}. 
The core width of the Hypatia and the mass parameter are free in the fits, the parameters which determine the transitioning points from the generalized hyperbolic resolution model to the exponential tails are constrained and the remaining parameters of the signal models are fixed. Constraining and fixing parameters in the signal model and other components is validated with simulation and pseudoexperiments to ensure unbiased parameter estimation and ensure valid coverage properties.
Combinatorial backgrounds are described by Chebychev polynomial functions up to order three; their coefficients allowed to float free in the fit. The order chosen for an individual fit is the one minimizing a likelihood that has been corrected for the number of degrees of freedom in the fit model, accounting for constraints and the number of parameters~\cite{Dauncey:2014xga}.
This likelihood is also used to select a baseline fit model among the various possibilities discussed in Sec.~\ref{sec:systematics}.

The only significant single-charm or charmless background contribution comes from the \mbox{$\LbLcKpiK$} decay. Its three-dimensional probability density function (PDF) is composed of a signal component in $m(\pKpi)$, a dedicated linear background in $m(\Kppim)$ or $m(\KKpim)$, and a conditional Gaussian PDF~in $m(\LcDzbK)$ or $m(\LcDs)$. The mean of the conditional Gaussian PDF depends linearly on $m(\Kppim)$ or $m(\KKpim)$, since a $\Dzb$ or $\Dsm$ mass constraint is used in the definition of $m(\LcDzbK)$ or $m(\LcDs)$.
In total, $84\pm13$ ($990\pm50$) \mbox{$\LbLcKpiK$} candidates remain in the  \mbox{$\LcDzbK$} (\mbox{$\LcDs$}) dataset used for the one-dimensional baseline fit. 
A subsequent one-dimensional fit, referred to as the baseline fit, which is independent from the three-dimensional fit, is used to determine the exclusive and partially reconstructed signal yields, that are needed for measuring the branching fractions.
In the baseline fit, the models of signal and combinatorial background are the same as in the three-dimensional fit, following the same strategy of floating, fixed and constrained parameters described earlier.
The yields of \mbox{$\LbLcKpiK$} decays are fixed in the baseline fit, and their uncertainties, including the anti-correlation with the signal yields, are propagated from the fit result of the three-dimensional fit to the statistical uncertainty of the \mbox{$\LbLcDzbK$} branching fraction.

The partially reconstructed quasi-two-body decays \mbox{$\LbLcDss$}, \mbox{$\LbLcDssz$} and \mbox{$\LbLcDsso$} are
described analytically. Their kinematic endpoints, which define the domain of the corresponding PDF, are fully determined by the masses of the decay products~\cite{PDG2022}. Each PDF is a superposition of Gaussian functions convolved with a two-sided step-function with a sloped plateau
(box-like), and an upward- or downward-open parabola~\cite{LHCb-PAPER-2021-052,LHCB-PAPER-2017-021}. These PDFs model the spin structure of the respective decays, and additionally describe a linear drop of efficiency
towards lower invariant masses, which is constrained to the value extracted from simulation.

In particular, the quasi-two-body decays through vector mesons, $\LbLcDssPRg$ and $\LbLcDssPRpi$ are modeled by parabolic shapes
facing downward and upward respectively, and due to the spin-$1/2$ initial state also involve a box-like component. Their relative normalization is constrained by the branching fraction of \mbox{$\mathcal{B}(\decay{\Dssm}{\Dsm\piz})$}.
Quasi-two-body decays through the spin-0 meson \mbox{$\decay{\Dsttstm}{\Dsm\piz}$}, and the spin-1 meson \mbox{$\decay{\Dstfsm}{\Dsm\g}$} are fully described by a box-like PDF. The latter is an effective description, since only about 20\% of the high mass tail of the \mbox{$\LbLcDssoPRg$} decay can be reconstructed in the chosen mass range, and variations of the spin structure in simulation do not significantly alter the shape.
The normalization of $\LbLcDsszPRpi$ and also $\LbLcDssoPRg$ decays is loosely constrained relative to that of the \mbox{$\LbLcDss$} component by means of the average of branching fraction measurements from the corresponding meson decays,
their simulated integral in the chosen invariant-mass range, and a correction accounting for the additional degrees of freedom in the spin structure of baryon decays~\cite{Datta:2003yk}.
Further decays to excited $\Dsm$ mesons are either kinematically forbidden or contribute at a significantly lower rate.

The shapes of partially reconstructed multibody decays, \ie \mbox{$\LbLcsDsPRpipi$} and partially reconstructed decays in \mbox{$m(\LcDzbK)$} cannot be modelled analytically unless their corresponding three-body dynamics is known and modeled.
Shapes of those decays are thus derived
from simulation in terms of a nonparametric PDF using a kernel density estimation method (KDE)~\cite{Cranmer:2000du}.
Fast simulation samples, either generated with \textsc{AmpGen}~\cite{AmpGen} or \textsc{RapidSim}~\cite{Cowan:2016tnm}, are used to model and cross-validate
the shapes of \mbox{$\LbLcsDsPRpipi$} and \mbox{$\LbScDzbK$} decays. As the shape of the partially reconstructed \mbox{$\LbLcDstzbK$} decay depends on the
three-body dynamics, a dedicated simulation sample including the most prominent $\Ds$ resonances, $\Dstftsm$ and $\Dstshm$,
as well as a small contribution from $\Xictsnz$ is used. The same sample is employed for the efficiency correction described in Sec.~\ref{sec:efficiencies}. The composition of simulated resonances is adapted to what is observed in the data. 
An effective correction for unconsidered three-body dynamics or efficiency effects
is obtained by multiplying the KDE template for the \mbox{$\LbLcDstzbK$} contribution with a first order polynomial function with a freely varying coefficient. That coefficient is anticorrelated ($-0.38$) to the leading coefficient of the polynomial describing the combinatorial background, and shared between the $\LbLcDstzbPRpiz$ and the $\LbLcDstzbPRg$ components. The normalization of those components is a free parameter in the fit, found to be consistent with the world-average value of $\BFRatioinline{\decay{\Dzb}{\Dzb\piz}}{\decay{\Dzb}{\Dzb\g}}$~\cite{PDG2022}.

\begin{figure}[ht]
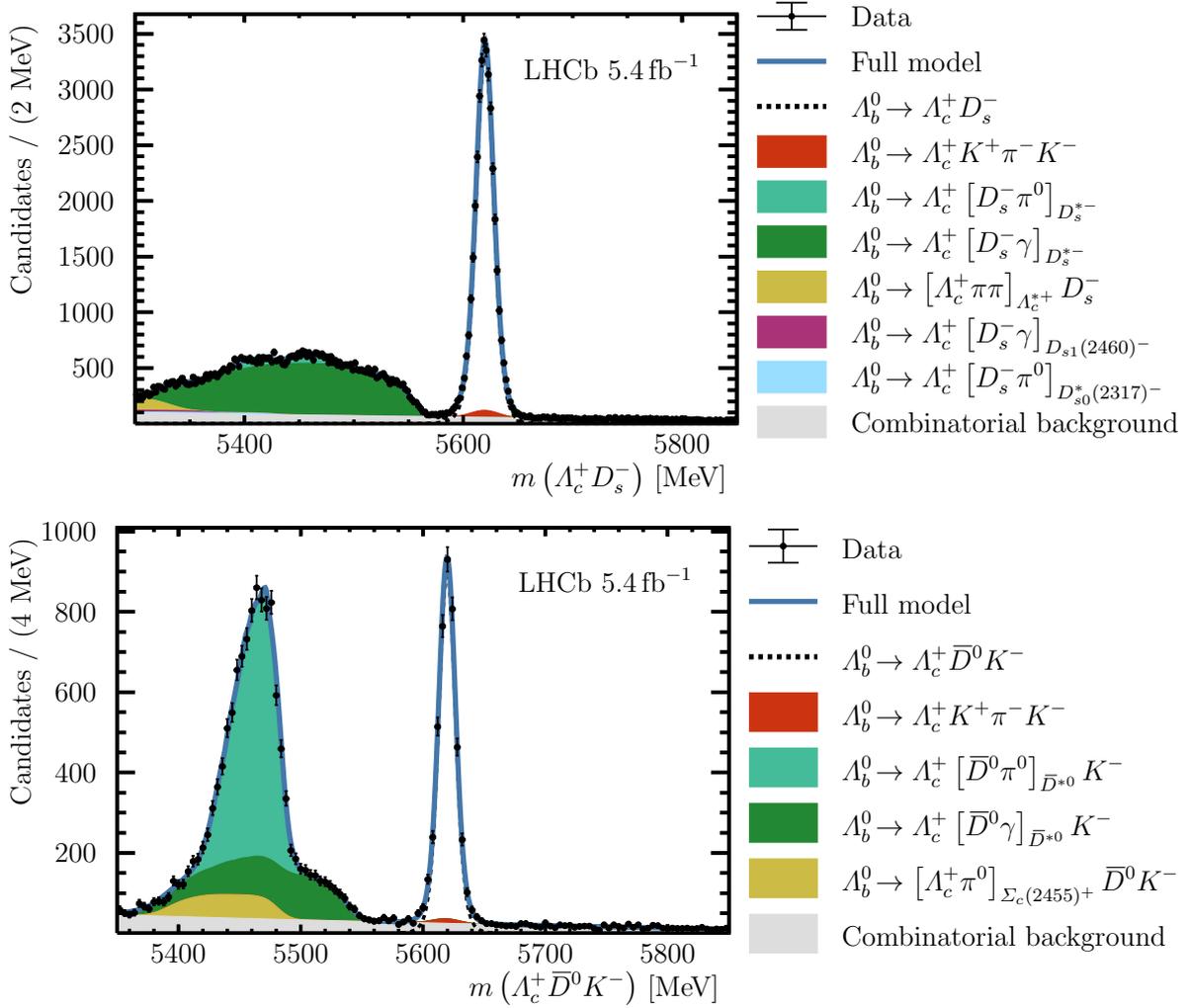

  \begin{adjustbox}{width=1\textwidth}
    \input{LcDsRun2_selA_fitA_multA_for_paper}
  \end{adjustbox}
  \begin{adjustbox}{width=1\textwidth}
    \input{LcD0KRun2_selA_fitA_multA_for_paper}
  \end{adjustbox}
  \caption[Fits]{Invariant-mass distributions of (top) \mbox{$\LcDs$} and (bottom) \mbox{$\LcDzbK$} candidates with the results of the baseline fit overlaid.}\label{fig:mass_fit}
\end{figure}

The baseline fits for the signal and normalization channel are shown in Fig.~\ref{fig:mass_fit}. The measured signal yields are
\begin{align*}
    N^{\LbLcDzbK} &= 4010\pm70, & N^{\LbLcDstzbK} &= 10\,560^{+310}_{-290}, \\
    N^{\LbLcDs} &= 35\,450 \pm 200, & N^{\LbLcDss} &= 46\,400 \pm 500,
\end{align*}
where the uncertainties are statistical, and the asymmetric uncertainties of the \mbox{$\LbLcDstzbK$} signal yield are driven by an anticorrelation with the \mbox{$\LbScDzbK$} component.

\section{Efficiency correction}\label{sec:efficiencies}
Efficiency ratios are calculated as the product of four factorizing terms:
\begin{itemize}
   \item Efficiencies of selections before the GEANT4 step of the simulation, referred to as generator level efficiencies hereafter.
   \item Trigger, reconstruction, and selection efficiencies are taken from simulation, up to but excluding the final selection step.
   \item Selection efficiencies of the final step are computed from simulation that has been weighted to match data.
   \item \mbox{$\LcpKpi$}~BDT efficiencies are evaluated using $\LbLcpi$ data.
\end{itemize}
Generator level efficiency ratios comprise the angular coverage of \lhcb, as well as loose kinematic and geometric selections of the generated candidates. Efficiencies are calculated only from those events for which a new underlying $pp$ interaction has been simulated.

The efficiency correction of the final selection step handles variables that are more difficult to model, like those that combine PID information~\cite{LHCb-DP-2018-001}.
These variables depend on detector multiplicity, production and decay kinematics, hereafter referred to as calibration variables.

Simulated events are weighted to match signal distributions in the phase-space of the calibration variables.
The set of 5 or 7 variables chosen for calibration are the track multiplicity, transverse momentum and \mbox{pseudorapidity} of the $\Lb$ candidate
for the production kinematics and the two square Dalitz variables of \mbox{$\LcpKpi$}, as well as those of \mbox{$\LbLcDzbstK$} for \mbox{$\Lc\Dzb\Km$} candidates.
It is not necessary to weight the simulated kinematics of the \mbox{$\DsKKpi$} decay in \mbox{$\LbLcDsorsm$}, due to the good agreement of the \evtgen model with the data~\cite{BaBar:2010wqe}.
The convention used for the square Dalitz variables follows that of Ref.~\cite{Back:2017zqt}:
\begin{equation*}
  m'\equiv \frac{1}{\pi} \arccos\left(2\frac{m_{12}-m_{12}^\mathrm{min}}{m^\mathrm{max}_{12}-m_{12}^\mathrm{min}}-1\right) \qquad \mathrm{and} \qquad
  \theta'\equiv \frac{1}{\pi} \theta_{12} \ ,
\end{equation*}
where $m^\mathrm{max}_{12}=m_P-m_{c_3}$ and $m_{12}^\mathrm{min}=m_{c_1}+m_{c_2}$ are the kinematic limits of $m_{12}$ in the $\decay{P}{c_1 c_2 c_3}$ decay,
while $\theta_{12}$ is the helicity angle between $c_1$ and $c_3$ in the rest frame of $c_1 c_2$. In the squared Dalitz plot of the $\decay{\Lc}{\proton\Km\pip}$, $P$ corresponds to $\Lc$, $c_1$ to the proton, $c_2$ to the kaon and $c_3$ to the pion; while $P$ corresponds to $\Lb$, $c_1$ to $\Lc$, $c_2$ to $\Dzb$ and $c_3$ to the companion kaon in the square Dalitz plot for the \mbox{$\LbLcDzbstK$}.

In the weighting, the 7 (5) dimensional calibration variable space is factorized into one 1D and three (two) 2D spaces for \mbox{$\LbLcDzbstK$}($\LbLcDsorsm$). This speeds up the weighting algorithm and allows for finer bins/partitions. For weighting, the Meerkat library~\cite{Poluektov:2014rxa} is used to create kernel density profiles of
calibration variable distributions from reconstructed simulation and signal data.
This method is validated against an adaptive binning algorithm, that split the calibration spaces in equally populated bins. Several configurations of the adaptive binning returned negligible differences with respect to the nominal kernel density method.
The assumption of factorization is verified with control plots that show that weighting in a certain subspace does not affect the other calibration variables. It is further validated by testing the weighting procedure with a method that does not factorize the space of calibration variables, but partitions the full phase space using gradient boosted decision trees~\cite{Rogozhnikov:2016bdp}.
 Without weighting, individual efficiencies are found to be larger across all studied decays, up to 2.5\%. In the ratio, the difference with respect to the weighted evaluation is between 0.6\% and 1.8\%.

The efficiency ratio for the \mbox{$\LcpKpi$}~BDT response is obtained in a data-driven manner using the $\LbLcpi$ calibration channel.
Efficiencies are obtained by fitting the calibration samples, weighted to match \mbox{$\LbLcDzbK$} or \mbox{$\LbLcDs$} data, simultaneously in ``pass'' and ``fail'' categories.
For validation tests, the requirements on the BDT response are equalized between signal and normalization channels.
The systematic uncertainty associated with the \mbox{$\LcpKpi$}~BDT efficiency correction is negligible due to the large size of the \mbox{$\LbLcpi$} samples, but also due to the similarity of selection requirements between this channel and the signal.

The overall efficiency ratios are found to be
\begin{align*}
    \epsilon^{\LbLcDzbK}/\epsilon^{\LbLcDs} &= 0.809\pm0.006, \\
    \epsilon^{\LbLcDstzbK}/\epsilon^{\LbLcDs} &= 0.689\pm0.005, \\
    \epsilon^{\LbLcDss}/\epsilon^{\LbLcDs} &= 0.785\pm0.005, 
\end{align*}
where the uncertainties are statistical.

\section{Systematic uncertainties}\label{sec:systematics}

\subsection{Invariant-mass fits}
As described in Sec~\ref{sec:fits}, the baseline model is the one that minimizes the likelihood corrected for the number of degrees of freedom in the fit model. Nine alternative models, returning comparable corrected likelihood values, are used in a discrete profiling method~\cite{Dauncey:2014xga}
to evaluate systematic uncertainties of the fit model.
Each of these alternative models concerns a single aspect of the baseline model. For example an exponential instead of a linear function is used as the combinatorial background description.
For each alternative, all other variations are tested in conjunction with the given variation, and the model that minimizes the corrected likelihood is taken
into consideration for the discrete profiling.
Changes to the baseline model are described below:
\begin{itemize}
  \item For signal, the baseline Gaussian-convolved Hypatia function is replaced with a double-sided Crystal Ball function~\cite{Skwarnicki:1986xj} with common mean.
        The double-sided Crystal Ball model improves the corrected likelihood in the \mbox{$\LcDzbK$} channel, but not in the normalization channel.
        For consistency, the Hypatia model is quoted as the baseline result, while the uncertainty calculation uses the alternative model as best fit
        to calculate the likelihood envelope of the discrete profiling method.
  \item Combinatorial background models are either Chebychev polynomials or exponential functions. It is found that either a first or second order polynomial function describes the combinatorial background best, even though higher-order functions can improve the uncorrected likelihood.
  \item In the baseline fit, the \mbox{$\LbLcDstzbK$} decays are modeled with a KDE approach, in which the $\Dstarzb$ decay modes are separated.
        An alternative model is a KDE template of the combined \mbox{$\LbLcDstzbK$} decay where the relative normalization of individual $\Dstarzb$ decay modes is fixed at the level of event generation.
  \item The $\LbScDzbK$ decays are modeled with KDE templates from \textsc{AmpGen} simulation, numerically convolved with a Gaussian function. One variation multiplies the shape with a first-order polynomial, to effectively correct for unconsidered three-body dynamics or efficiency effects, another one adds contributions from $\Sctftp$ decays. When allowing the normalization of this component to float free in the fit, it is compatible with zero yield. A model is employed, that loosely constrains the ratio of yields of the components modeling the decays through $\Sctftp$ and $\Sctfffp$ states. The value of the constraint is based on the approximate ratios observed in $\decay{\Lb}{\Lc\Km\Kp\pim}$~\cite{LHCB-PAPER-2020-028} and $\decay{\Lb}{\Lc\pim\pip\pim}$~\cite{LHCB-PAPER-2011-016} decays.
  \item The normalization fraction of box-like shapes with respect to parabolic shapes are separated for the \mbox{$\LbLcDssPRpi$} and \mbox{$\LbLcDssPRg$} components.
        In addition, the slope parameter that models the efficiency gradient and the $\Dssm$ branching fractions is fixed, instead of constrained, to values obtained from simulation.
  \item The $\LbLcDspipi$ decays are effectively modeled by a KDE template derived from an \textsc{AmpGen} simulation sample where the decay occurs through the $\Lctstfp$ resonance.
        An alternative model is the decay through the $\Lctfnfp$ state.
\end{itemize}

To compute the envelope for the discrete profiling method analytically, likelihoods are approximated with
bifurcated parabolas, which account for asymmetric uncertainties.
From this method, the yield ratios obtained, including their statistical and systematic uncertainties due to the fit model, are
\begin{align*}
  \begin{split}
    N^{\LbLcDzbK}/N^{\LbLcDs}   & = 0.1132 {}_{-0.0020}^{+0.0021} {}_{-0.0007}^{+0.0006}, \\
    N^{\LbLcDstzbK}/N^{\LbLcDs} & = 0.298 {}_{-0.008}^{+0.009} {}_{-0.009}^{+0.008}, \\
    N^{\LbLcDss}/N^{\LbLcDs}    & = 1.309 \pm 0.017 {}_{-0.043}^{+0.047}\ .
  \end{split}
\end{align*}
The dominating systematic effect comes from the signal shape variation in the determination of the exclusive $\LbLcDs$ and $\LbLcDzbK$ yields. For the measurement of $N^{\LbLcDstzbK}$, the dominant source of uncertainty is the multiplication of the $\LbScDzbK$ component by a first-order polynomial function. Changing the combinatorial background description, and separating the normalization fraction of the contributions in the description of \mbox{$\LbLcDss$} decays, are the largest contributions to the uncertainty on $N^{\LbLcDss}$.

\subsection{Parameters of the simulation weighting}
Efficiency ratios are calculated using five different settings of the initial phase-space binning to generate the adaptive KDE profiles.
The results are found to be very stable against changing this parameter from the default 80 bins to 20, 40, 160 and 320 bins.
Their standard deviation with respect to the baseline result is taken as the systematic uncertainty.

\subsection{Multiple candidates} \label{sec:multsyst}
Multiple candidates are randomly removed from the sample, as they are mainly composed of candidates where a signal track has been swapped with a combinatorial track that happens to have similar kinematics.
In the case of \mbox{$\LbLcDzbstK$}, another source of multiple candidates exists, swapping the companion $\Km$ with that from the $\Lc$ decay.
Removing this background entirely is inefficient, but since the fraction of multiple candidates in the \mbox{$\LcDzbK$} channel is larger (2.4\% compared to 1.1\% in \LcDs), further methods to remove multiple candidates are studied: namely using the minimal $\chi^2/$ndf of a kinematic fit of the decay chain~\cite{Hulsbergen:2005pu}, or the maximum sum of PID variables to select the best out of the multiple candidates. The resulting maximum deviation from the baseline result is small, but assigned as systematic uncertainty.

\subsection{Simulation and control sample sizes}
Statistical uncertainties of the generator level efficiency ratios, and simulation and calibration sample sizes are propagated to the branching ratio measurement, as summarized in Table~\ref{tab:sys}.
For generator level efficiencies, only those events for which a new underlying $pp$ interaction has been simulated can be taken into account for the uncertainty calculation.
As the underlying $pp$ interaction is reused 100 times~\cite{LHCb-DP-2018-004}, with an independently
generated signal decay for each simulated event, the uncertainty on the generator level efficiency ratio is treated independently of the statistical uncertainty of the final simulation sample size.

\subsection{Summary}\label{sec:sys_sum}
Table~\ref{tab:sys} summarizes the systematic uncertainties for the three measured ratios of branching fractions.
\begin{table}[tb]
  \caption[Systematic uncertainties]{Systematic uncertainties relative to the branching fraction ratio measurements. The relative statistical uncertainty is shown as a reference. Values are given in percent.}
\begin{tabular}{l c c c}
  \toprule
  \multirow{2}{*}{Source / relative to} & \footnotesize{$\BFRatio{\LbLcDzbK}{\LbLcDs}$} &  \footnotesize{$\BFRatio{\LbLcDstzbK}{\LbLcDs}$} & \footnotesize{$\BFRatio{\LbLcDss}{\LbLcDs}$}\\
   & \multicolumn{1}{c}{[\%]} & \multicolumn{1}{c}{[\%]} & \multicolumn{1}{c}{[\%]} \\
  \midrule
  Fit model              & ${}^{+0.5}_{-0.6}$
                         & ${}^{+2.8}_{-3.0}$
                         & ${}^{+3.6}_{-3.3}$\\
  Weighting              & $0.1$ & $0.1$ & $0.0$ \\
  Multiple candidates    & $0.0$ & $0.0$
                         & $0.1$ \\
  Size of the simulated samples       & $0.4$ & $0.3$ & $0.2$ \\
  Size of the generated samples  & $0.6$ & $0.6$ & $0.6$ \\
  \midrule  
  Total                  & $0.9$
                         & ${}^{+2.9}_{-3.1}$
                         & ${}^{+3.7}_{-3.3}$  \\
  \midrule
  Statistical            & 1.8 & 2.8 & 1.3 \\
  \bottomrule
\end{tabular}\label{tab:sys}
\end{table}
The choice of the fit model is found to dominate the uncertainty for partially reconstructed decays, while the systematic uncertainties  for \mbox{$\LbLcDzbK$} are small compared to their corresponding statistical uncertainty.

\section{Conclusion}\label{sec:results}
The ratio of branching fractions for the \mbox{$\LbLcDzbK$}, \mbox{$\LbLcDstzbK$} and \mbox{$\LbLcDss$} decays, relative to that of the \mbox{$\LbLcDs$} decay, are measured in $\proton\proton$ collisions
at $\sqs=13\tev$ corresponding to an integrated luminosity of 5.4\invfb collected with the \lhcb detector.
The results are found to be
\begin{align*}
  \begin{split}
    \BFRatio{\LbLcDzbK}{\LbLcDs}\cdot\BFRatio{\DzbKpi}{\DsmKKpi} & = 0.1400 {}_{-0.0025}^{+0.0026} {}_{-0.0013}^{+0.0012}, \\
    \BFRatio{\LbLcDstzbK}{\LbLcDs}\cdot\BFRatio{\DzbKpi}{\DsmKKpi} & = 0.432 {}_{-0.012}^{+0.013} \pm 0.013,
  \end{split}
\end{align*}
where the first uncertainties are statistical and the second systematic.
In the $\LcDzbK$ channel, the ratio of branching fractions of $\Lb$ decays proceeding through the excited compared to the ground state $\Dzb$ is measured to be
\begin{equation*}
    \BFRatio{\LbLcDstzbK}{\LbLcDzbK}=3.09^{+0.11}_{-0.10}{}^{+0.09}_{-0.10},
\end{equation*}
where correlations between uncertainties are taken into account, but are found to be small.
Including the known values of the $\D$ meson branching fractions from Ref.~\cite{PDG2022}, the ratios of branching fractions are
\begin{align*}
  \begin{split}
    \BFRatio{\LbLcDzbK}{\LbLcDs}   & = 0.1908 {}_{-0.0034}^{+0.0036} {}_{-0.0018}^{+0.0016} \pm 0.0038, \\
    \BFRatio{\LbLcDstzbK}{\LbLcDs} & = 0.589 {}_{-0.017}^{+0.018} {}_{-0.018}^{+0.017} \pm 0.012, \\
    \BFRatio{\LbLcDss}{\LbLcDs}    & = 1.668 \pm 0.022 {}_{-0.055}^{+0.061},
  \end{split}
\end{align*}
where the third uncertainties are due to the uncertainty of the branching fractions of $\DsmKKpi$ and $\DzbKpi$ decays.

The result obtained for $\BFRatio{\LbLcDss}{\LbLcDs}$ is compatible with several predictions~\cite{Giri:1997te,Fayyazuddin:1998ap,Mohanta:1998iu,Zhao:2018zcb,Gutsche:2018utw,Ke:2019smy,Chua:2019yqh}.

To probe factorization approaches in the \mbox{$\LbLcDzbstK$} decays,
the following values for the double ratios $\mathcal{DR}^{(*)}$, defined in Eq.~\eqref{eq:dr}, are obtained
\begin{align*}
    \mathcal{DR}(\Bzb)&=1.29\pm0.20, & \mathcal{DR}^{*}(\Bzb)&=1.28\pm0.19,\\
    \mathcal{DR}(\Bm)&=1.20\pm0.30, & \mathcal{DR}^{*}(\Bm)&=0.87\pm0.12,\\
    \mathcal{DR}(\Bcm)&=1.3\pm0.5, & \mathcal{DR}^{*}(\Bcm)&=0.8\pm0.4,
\end{align*}
assuming uncorrelated uncertainties, and taking known values for the mesonic branching fractions from Ref.~\cite{PDG2022}.
Larger baryonic branching fractions are expected, because of an additional color-suppressed amplitude (see Fig.~\ref{fig:feynman}) in the $\Lb$ decay, which does not exist for mesons, however the measured ratios are still inconclusive.

The ratios of branching fractions that are relevant for pentaquark searches, see Eq.~\eqref{eq:pcbf}, are
\begin{align*}
  \begin{split}
    \BFRatio{\decay{\Lb}{\jpsi\proton\Km}}{\LbLcDzbK}&=0.152^{+0.032}_{-0.028}, \\
    \BFRatio{\decay{\Lb}{\jpsi\proton\Km}}{\LbLcDstzbK}&=0.049^{+0.011}_{-0.009},
  \end{split}
\end{align*}
using results from Ref.~\cite{LHCb-PAPER-2015-032}, and assuming that uncertainties are uncorrelated.
A future search for pentaquarks in the \mbox{$\LbLcDzbstK$} decays will be able to determine the fit fractions $f_{\Lc\DzborDstarzb}(\Pcp)$, which can be used to test model predictions of the ratio of $\Pcp$ branching fractions \mbox{$\mathcal{B}(\decay{\Pcp}{\Lc\DzborDstarzb})/\mathcal{B}(\decay{\Pcp}{\jpsi\proton})$}.

\section*{Acknowledgements}
%
%
\noindent We express our gratitude to our colleagues in the CERN
accelerator departments for the excellent performance of the LHC. We
thank the technical and administrative staff at the LHCb
institutes.
We acknowledge support from CERN and from the national agencies:
CAPES, CNPq, FAPERJ and FINEP (Brazil); 
MOST and NSFC (China); 
CNRS/IN2P3 (France); 
BMBF, DFG and MPG (Germany); 
INFN (Italy); 
NWO (Netherlands); 
MNiSW and NCN (Poland); 
MCID/IFA (Romania); 
MICINN (Spain); 
SNSF and SER (Switzerland); 
NASU (Ukraine); 
STFC (United Kingdom); 
DOE NP and NSF (USA).
We acknowledge the computing resources that are provided by CERN, IN2P3
(France), KIT and DESY (Germany), INFN (Italy), SURF (Netherlands),
PIC (Spain), GridPP (United Kingdom), 
CSCS (Switzerland), IFIN-HH (Romania), CBPF (Brazil),
and Polish WLCG (Poland).
We are indebted to the communities behind the multiple open-source
software packages on which we depend.
Individual groups or members have received support from
ARC and ARDC (Australia);
Key Research Program of Frontier Sciences of CAS, CAS PIFI, CAS CCEPP, 
Fundamental Research Funds for the Central Universities, 
and Sci. \& Tech. Program of Guangzhou (China);
Minciencias (Colombia);
EPLANET, Marie Sk\l{}odowska-Curie Actions, ERC and NextGenerationEU (European Union);
A*MIDEX, ANR, IPhU and Labex P2IO, and R\'{e}gion Auvergne-Rh\^{o}ne-Alpes (France);
AvH Foundation (Germany);
ICSC (Italy); 
GVA, XuntaGal, GENCAT, Inditex, InTalent and Prog.~Atracci\'on Talento, CM (Spain);
SRC (Sweden);
the Leverhulme Trust, the Royal Society
 and UKRI (United Kingdom).

\addcontentsline{toc}{section}{References}
\bibliographystyle{LHCb}
\bibliography{main,standard,LHCb-PAPER,LHCb-CONF,LHCb-DP,LHCb-TDR}

\newpage
\centerline
{\large\bf LHCb collaboration}
\begin
{flushleft}
\small
R.~Aaij$^{35}$\lhcborcid{0000-0003-0533-1952},
A.S.W.~Abdelmotteleb$^{54}$\lhcborcid{0000-0001-7905-0542},
C.~Abellan~Beteta$^{48}$,
F.~Abudin{\'e}n$^{54}$\lhcborcid{0000-0002-6737-3528},
T.~Ackernley$^{58}$\lhcborcid{0000-0002-5951-3498},
B.~Adeva$^{44}$\lhcborcid{0000-0001-9756-3712},
M.~Adinolfi$^{52}$\lhcborcid{0000-0002-1326-1264},
P.~Adlarson$^{78}$\lhcborcid{0000-0001-6280-3851},
C.~Agapopoulou$^{46}$\lhcborcid{0000-0002-2368-0147},
C.A.~Aidala$^{79}$\lhcborcid{0000-0001-9540-4988}
% ...

{\footnotesize \it

% ...
$^{35}$Nikhef National Institute for Subatomic Physics, Amsterdam, Netherlands\\
% ...
$^{44}$Instituto Galego de F{\'\i}sica de Altas Enerx{\'\i}as (IGFAE), Universidade de Santiago de Compostela, Santiago de Compostela, Spain\\
% ...
$^{46}$European Organization for Nuclear Research (CERN), Geneva, Switzerland\\
% ...
$^{48}$Physik-Institut, Universit{\"a}t Z{\"u}rich, Z{\"u}rich, Switzerland\\
% ...
$^{52}$H.H. Wills Physics Laboratory, University of Bristol, Bristol, United Kingdom\\
% ...
$^{54}$Department of Physics, University of Warwick, Coventry, United Kingdom\\
% ...
$^{57}$School of Physics and Astronomy, University of Glasgow, Glasgow, United Kingdom\\
$^{58}$Oliver Lodge Laboratory, University of Liverpool, Liverpool, United Kingdom\\
% ...
$^{66}$Syracuse University, Syracuse, NY, United States\\
% ...
$^{78}$Department of Physics and Astronomy, Uppsala University, Uppsala, Sweden, associated to $^{57}$\\
$^{79}$University of Michigan, Ann Arbor, MI, United States, associated to $^{66}$\\
% ...
}
\end{flushleft}
\end{document}